# A counter-example to Bell's theorem with a 'softened' singularity and a critical remark to the implicit demand that physical signals may not travel faster than light.

Author: J.F. Geurdes
Address:   C. van der Lijnstraat 164
           2593 NN Den Haag
Email:     hgeurdes@flashmail.com

**Abstract**

In the present paper a counter-example to Bell's theorem is given which is based on common probability densities as standard normal (Gaussian) and uniform probability densities. The reason for violating the Bell inequalities lies in the 'softening' of functions similar to the Dirac delta such that they can be 'hidden' inside a sign function.

## I. Introduction.

It is commonly known that statistics and probability theory are widely applied in physics. Perhaps it is not so well known that, in fundamental quantum physics, there exists a statistical problem which appears difficult, if not impossible, to solve with Kolmogorovian (McCord and Moroney (1964), Hogg and Graig (1970), Mood, Graybil and Boes, (1974)) probability theory. In the present paper a solution to this problem is advanced. The origin of the problem lies in the Bell inequalities.

Bell inequalities, originally derived by Bell (1964) and later developed by, for instance, Greenberger (1995) and Mermin (1995), are extremely important to the statistics of experiments in fundamental physics. With Bell inequalities, Einstein's doubts about the completeness of quantum mechanics, Einstein, Rosen and Podolsky (1935), could be studied experimentally. A decisive experiment, based on Bell inequalities, was performed by Aspect, Dalibard and Roger (1982).

The original argument of Einstein, Podolsky and Rosen (EPR), which voiced some of Einstein's doubts about the completeness of quantum mechanics, is based on three premises. The first one is the doctrine of realism, that is, the observed phenomena are caused by physical realities whose existence is independent of human observers. The second one expresses faith in inductive inferential reasoning. The third one states that there are no influences possible that travel faster than light.

The Bell inequalities have unearthed a tremendous amount of research in the physics and statistics of experiments. Recent progress has been towards the necessity of Bell inequalities, studied by, for instance, Jordan (1994), loophole-free tests of Bell inequalities, studied by Fry, Walther and Li (1995) and continuous variables, studied by Tara and Agarwal (1994). In addition, Scully and Cohen (1986) studied the relation between the Wigner distribution function and the EPR problem, Jarrett (1986) examined time dependent hidden states and Pitowsky (1983) investigated the possibility of deterministic models.

Recent progress has also been towards a change in probability theory. An interesting change is to introduce complex numbers into probability theory. Studies in this particular field are performed by, for instance, Youssef (1995) and Gudder (1993).

The present author will argue below, however, that such a radical change in probability theory is unnecessary. Moreover, it is noted that the author (Geurdes (1998 a, b), Geurdes (2001)) already has demonstrated that, e.g. complex probability spaces are unneccesary to reduplicate quantum mechanical results from local hidden variables models.

For completeness a short description of the principles of a typical Bell inequality experiment will be presented. The physical situation of the Bell inequalities refer to a correlation between spin states of spatially separated particles, originally in the singlet state and arising from a single source. Restricting oneself to electrons, the spin of the electron can be seen as the intrinsic magnetic moment of the electron, while the singlet state means that both electrons arose from the source with opposite spins. Simply stated, the EPR paradox demonstrates that electrons maintain a correlation between their respective spin variables, despite of a great, theoretically infinite, spatial separation. This type of experiment is a simplification of the original 'Gedanken' experiment of Einstein, Podolsky and Rosen.

The correlation function, P($\underline{a},\underline{b}$), representing a hidden variable explanation of the observed correlation between, spatially separated, spin measurements, is given by

$$P(\underline{a},\underline{b}) = \int d\underline{\lambda}\, \rho(\underline{\lambda}) A(\underline{\lambda},\underline{a}) B(\underline{\lambda},\underline{b}). \qquad 1$$

Here, $\underline{a}=(a_1,a_2,a_3)$ and $\underline{b}=(b_1,b_2,b_3)$ are, unitary, parameter vectors of the measurement functions $A(\underline{a},\lambda)=\pm 1$ and $B(\underline{b},\lambda)=\pm 1$, which are also supposed to depend on the (set of) hidden variable(s) $\lambda$. Here, $A=+1$, represents, for instance, the discrete outcome 'spin-up' at the A apparatus, while, $A=-1$, represents the discrete outcome 'spin-down'. The parameter vectors represent the spatial orientation of the measurement apparatuses.

Because measurement, A, does not depend on parameter vector, $\underline{b}$, while measurement, B, does not depend on parameter vector, $\underline{a}$, so called Einstein locality is warranted. This means that the real factual situation of system $S_2$ is independent of what is done with system $S_1$, which is spatially separated from the former, Einstein (1949).

In equation (1), the function, $\rho=\rho(\lambda)$, represents the probability density of the local hidden variables, $\lambda$, while the integration is performed over the whole range of hidden variables. Of course, $\rho=\rho(\lambda)$, is also independent of the parameter vectors, $\underline{a}$, and, $\underline{b}$.

For the correlation function in expression (1), the following inequality

$$|P(\underline{a},\underline{b}) - P(\underline{a},\underline{d})| + P(\underline{b},\underline{c}) + P(\underline{c},\underline{d}) \le 2, \qquad 2$$

can be derived. Subsequently it can be verified that the quantum correlation, $P_{qm}(\underline{a},\underline{b})=-(a_1 b_1 + a_2 b_2 + a_3 b_3)$, violates this inequality. Hence, it was generally concluded that the local hidden variable correlation cannot be equal to the quantum correlation. Moreover, if in experiment, the inequality is violated, the quantum correlation is, most probably, the correct description of the correlation between spatially separated particles.

At first instance, the previous argument against local hidden variables appear solid. However, it should be noted that the absolute exclusion of hidden variable models, depends on the universal validity of the inequality. The total exclusion of local hidden variable models implies that the quantum correlation cannot be reproduced by any model within basic probability theory. However, this has not been conclusively proved yet. The only thing that has been demonstrated is that the quantum correlation contradicts the inequality. This may appear sufficient evidence against all local hidden variables. The scope of the evidence can be questioned nonetheless.

In section II, a classical probability model is presented that is the probabilistic basis of the model. In section III, it is explained how to 'soften' a Dirac delta function such that it can occur in a sign function and, hence, be included in the A and B measurement functions. This part is an important section of the paper because it points to the reason why Bell's theorem is incomplete. In section IV, the model functions are evaluated. In section V, the result is discussed and an alternative way to introduce a singular function, Dirac delta type of function, into a sign function is discussed also.

*II. Probability Model.*

The proposed model employs a classical probability density which has three components. If we write, $\rho_{tot}=\rho_{Norm}\rho_I\rho_{II}$, then it is intended to have

$$\rho_{Norm}(c_1,c_2,c_3) = \frac{1}{(2\pi)^{3/2}} \exp\left[-\frac{1}{2}\sum_{k=1}^{3} c_k^2\right], \qquad 3$$

where the variables, $\{\chi_k\}_{k=1,2,3}$ are shared by the two particles. In addition, the $\rho_I$, is a density for local variables that reside only on one particle. We have,

$$r_I = \frac{(1+m_I)}{32\,T(e)}[d(1+m_I)+d(1-m_I)]q[\frac{1}{4}+x_I]q[\frac{1}{4}-x_I]$$

$$\times \prod_{k=1}^{3} q[1+m_{kI}]q[1-m_{kI}]q[T(n)+t_I]q[T(n)-t_I]$$

and a similar expression for $\rho_{II}$

$$r_{II} = \frac{(1+m_{II})}{32\,T(e)}[d(1+m_{II})+d(1-m_{II})]q[\frac{1}{4}+x_{II}]q[\frac{1}{4}-x_{II}]$$

$$\times \prod_{k=1}^{3} q[1+m_{kII}]q[1-m_{kII}]q[T(n)+t_{II}]q[T(n)-t_{II}]$$

Here, $T(n)=n\_0$. Note the difference between on the one hand, $n_I$ and $n_{II}$ and 'plain' n.

The associated integration procedure contains three parts. The first part, in bracket notation, is the integration over $\rho_I$-type of variables. We have

$$(f)_I = \int dm_I \sum_{n_I=0}^{1} \int dx_I \prod_{k=1}^{3} \int d\,m_{kI} \int d\,t_I\; r_I\; f$$

where for all variables there is an integration from minus infinity to plus infinity. Secondly for $\rho_{II}$-type of variables

$$(f)_{II} = \int dm_{II} \sum_{n_{II}=0}^{1} \int dx_{II} \prod_{k=1}^{3} \int d\,m_{kII} \int d\,t_{II}\; r_{II}\; f$$

Hence, it easily follows that, $(1)_I=(1)_{II}=1$. In addition, because in $\rho_{tot}$, only 'ordinary' partial probability densities are employed, we have the integration

$$( f )_N = \int_{-\infty}^{\infty} \int_{-\infty}^{\infty} \int_{-\infty}^{\infty} d\mathbf{c}_1 d\mathbf{c}_2 d\mathbf{c}_3 \, f \, r_{Norm} \qquad 8$$

Hence, genuine probabilities arise from, $\rho_{Norm}$, and consequently, genuine probabilities arise from, $\rho_{tot}$. Moreover, the complete integration of a proper function F times the density $\rho_{tot}$ is defined by, $(F)=(((F)_I)_{II})_N$. Concludingly, a probability density containing standard normal and uniform densities is employed in the probability part of the model.

### III. Measurement equations.

Subsequently, the measurement functions A and B are introduced. Beforehand let us note that A is independent of the $b_k$, k=1,2,3, unity, parameter vector for B, while, B is independent for, $a_k$, k=1,2,3, the unity parameter vector for A. Hence, locality in the sense of Einstein is maintained in the measurement model. The correlation in the model consequently must follow from the postulated local hidden variables, as expressed in the probability density function.

Given the expressions,

$$\mathbf{s}_a = \sum_{k=1}^{3} a_k \, sign \, \mathbf{c}_k$$

$$\qquad 9$$

$$\mathbf{s}_b = \sum_{k=1}^{3} b_k \, sign \, \mathbf{c}_k ,$$

the measurement functions A and B can be expressed by

$$A = \sum_{k=1}^{3} \mathbf{i}_k(\mathbf{s}_a) \, s_{kI}(\mathbf{s}_a)$$

$$\qquad 10$$

$$B = -\sum_{k=1}^{3} \mathbf{i}_k(\mathbf{s}_b) \, s_{kII}(\mathbf{s}_b)$$

Here the functions, $\iota_k(\sigma_a)$, are defined by, $\iota_k(\sigma_a)=\iota[\sigma_a \in I_k]$, with,

$$\mathbf{i}[x \in Y] = 1 \_ x \in Y$$

$$\qquad 11$$

$$\mathbf{i}[x \in Y] = 0 \_ x \notin Y.$$

The sets, $I_k$, in the A and B are given by

$I_1 = \{x \in R \,|\, -\sqrt{3} \leq x < -1\}$,

$I_2 = \{x \in R \,|\, -1 \leq x \leq 1\}$, 12

$I_3 = \{x \in R \,|\, 1 < x \leq \sqrt{3}\}$.

The ι functions are employed to single-out the specific intervals for the σ. Moreover, the $s_{kI}$ and $s_{kII}$ factors in A are given by

$$s_{1I} = [\, n_I \; sign[\, \sigma_a + 1 - m_{1I}\,] - d_{0,n_I}\,] sign\,[\,T\Delta_n(\,f_I(\,x_I\,)\,) - \tau_I\,]\,, \qquad 13$$

with, $f_I(x_I) = x_I^2 - (1/n)^2$. note that when, $\iota_1=1$, and hence, $\iota_2=\iota_3=0$, the $\sigma_a+1$ is in the interval [-1,1], because as can be verified, $|\sigma_a| \leq 3^{1/2}$, as well as, $|\sigma_b| \leq 3^{1/2}$. In addition, $s_{3I}$, is written by

$$s_{3I} = [\, n_I \; sign[\, \sigma_a - 1 - m_{3I}\,] + d_{0,n_I}\,] sign\,[\,T\Delta_n(\,f_I(\,x_I\,)\,) - \tau_I\,]\,. \qquad 14$$

Similarly, when, $\iota_3=1$, hence, $\iota_1=\iota_2=0$, then, $\sigma_a-1$ is in the interval [-1,1].

In both cases, the factor, $\Delta_n(f_I(x_I))$, is closely related to the Dirac delta (Levoine, 1963, Lighthill 1958) function. We have

$$\Delta_n(\,f_I(\,x_I\,)) = \frac{2}{p}\frac{1}{1+n^2\,f_I^2(\,x_I\,)}. \qquad 15$$

As can be seen easily, $0 < T(n)\Delta_n < n$. Hence, $\Delta_n$, can be 'hidden' inside a sign-function with a 'counter variable', like $\tau_I$, ranging from -n to +n, where, n, approaches infinity.

Furthermore, we see that integration leads to

$$\lim_{n\to\infty}\int_{x=-\frac{1}{4}}^{x=\frac{1}{4}}\Delta_n(\,x^2 - (1/n)^2\,)\,dx = \lim_{n\to\infty}\frac{1}{n}\int_{x=-\frac{1}{4}}^{x=\frac{1}{4}}\frac{2\,n}{p}\frac{dx}{1+n^2\,(\,x^2-(1/n)^2\,)^2} \qquad 16$$

Hence, substitution of $y = x_I^2 - (1/n)^2$, gives, $dy = 2x\,dx$, hence,

$$\lim_{n\to\infty} \int_{x=-\frac{1}{4}}^{x=\frac{1}{4}} \Delta_n(x^2-(1/n)^2)\, dx = \lim_{n\to\infty} \frac{-1}{n}\int_{x=-\frac{1}{4}}^{x=0} \frac{2n}{p}\frac{dy}{1+n^2 y^2}\frac{1}{2\sqrt{y+(1/n)^2}}$$

17

$$+\lim_{n\to\infty} \frac{1}{n}\int_{x=0}^{x=\frac{1}{4}} \frac{2n}{p}\frac{dy}{1+n^2 y^2}\frac{1}{2\sqrt{y+(1/n)^2}}.$$

Because, in the limit n to infinity, only points very close around y=0 give a non-zero integrand, we may write

$$\lim_{n\to\infty} \int_{x=-\frac{1}{4}}^{x=\frac{1}{4}} \Delta_n(x^2-(1/n)^2)\, dx = \lim_{n\to\infty} \frac{-1/n}{1/n}\int_{x=-\frac{1}{4}}^{x=0} \frac{n}{p}\frac{dy}{1+n^2 y^2}$$

18

$$+\lim_{n\to\infty} \frac{1/n}{1/n}\int_{x=0}^{x=\frac{1}{4}} \frac{n}{p}\frac{dy}{1+n^2 y^2}.$$

Hence, the result of integration gives

$$\lim_{n\to\infty} \int_{x=-\frac{1}{4}}^{x=\frac{1}{4}} \Delta_n(x^2-(1/n)^2)\, dx =$$

19

$$\lim_{n\to\infty}\{-\frac{1}{p}\{arctg[n(x^2-(1/n)^2)]\}_{x=-1/4}^{x=0}+\frac{1}{p}\{arctg[n(x^2-(1/n)^2)]\}_{x=0}^{x=1/4}\} = 1$$

If in the integration over the $x_I$ and $x_{II}$ type of variables there is no sign over the difference of the $\Delta$ and the tau, then 1/2, results, while, if the sign is present, we find that, unity is the result. This will be explained more carefully in the next section.

Alternatively we also find from the expression

$$d(y) = \lim_{n \to \infty} \frac{n}{p} \frac{1}{1+n^2 y^2},$$

that it is possible to obtain

$$\lim_{n \to \infty} \int_{y=-1/n^2}^{y=\frac{1}{16}-1/n^2} \frac{2n}{p} \frac{dy}{1+n^2 y^2} \frac{1}{2n\sqrt{y+(1/n)^2}} = \int_{y=0}^{y=1/16} d(y) \lim_{n \to \infty} \frac{1}{n\sqrt{y+1/n^2}} dy$$

$$= \frac{1}{2} \lim_{n \to \infty} \frac{1/n}{\sqrt{1/n^2}} = \frac{1}{2}.$$

Similarly for the negative branch of the integral,

$$\lim_{n \to \infty} \int_{y=\frac{1}{16}-1/n^2}^{-1/n^2} \frac{-2n}{p} \frac{dy}{1+n^2 y^2} \frac{1}{2n\sqrt{y+(1/n)^2}} = \int_{y=0}^{y=1/16} d(y) \lim_{n \to \infty} \frac{1}{n\sqrt{y+1/n^2}} dy$$

$$= \frac{1}{2} \lim_{n \to \infty} \frac{1/n}{\sqrt{1/n^2}} = \frac{1}{2}.$$

Hence, we also find that

$$\lim_{n \to \infty} \int_{x=-\frac{1}{4}}^{x=\frac{1}{4}} \Delta_n(x^2 - (1/n)^2) dx = \frac{1}{2} + \frac{1}{2} = 1.$$

A second alternative to the present delta will be discussed below.

In this sense, the singularity can be 'softened' to be hidden inside a sign function.

In addition, the $s_{2I}$, is defined by

$$s_{2I} = sign[\, s_a - m_{2I}\, ].  \qquad 20$$

This completes the definition of the A-wing measurement function. The B-wing measurement function is defined similarly, as can be seen from eq. (10). Only, in the definition of each term, each occurrence, $\sigma_a$, is replaced by, $\sigma_b$, while all indices, 'I' are replaced by indices, 'II'.

*IV Evaluations.*

In the evaluation of the different integrals over the combined model, we first note that for the normal density it follows that

$$\int_{-\infty}^{\infty}\int_{-\infty}^{\infty}\int_{-\infty}^{\infty} d\,c_1\, d\,c_2\, d\,c_3\, r_{Norm}\, sign\, c_i\, sign\, c_j = d_{i,j} \qquad 21$$

Secondly, when p in the interval [-1,1], integration over uniform distributed variable, $\mu$, leads to

$$\frac{1}{2}\int_{-1}^{1} sign(p - m)\, d\,m = p. \qquad 22$$

This type of evaluation will be needed very often. A special case to this is the integration over $\tau_I$ and/or $\tau_{II}$. We then have,

$$\frac{1}{2T}\int_{-T}^{T} sign[T\Delta_n - t\,]\, d\,t = \Delta_n. \qquad 23$$

Thirdly, it may be noted that

$$\int_{-\infty}^{\infty} [\,\boldsymbol{d}(1+m) + \boldsymbol{d}(1-m)\,](1+m)\,dm = \sum_{m \in \{-1,1\}} (1+m) = 2. \qquad 24$$

After having presented the basic needs for evaluating the model, let us turn to the evaluation of the mean (A). First we evaluate $(A)_I$. Here, we note that we need to evaluate the three s-functions, because, the $\iota_j$, $j=1,2,3$, refer to exclusive sets. Hence, $(s_{1I})_I$, is evaluated. Because, $\sigma_a+1$ is in the interval $[-1,1]$ in this case, the evaluation over $\mu_{1I}$, results into the term $\{n_I(\sigma_a+1)-\delta_{0,n_I}\}$. Observing eq.(22) and the approximation in eq. (16), this subsequently leads to

$$(s_{1I})_I = \sum_{m_I} \frac{1+m_I}{8} \sum_{n_I=0}^{1} \frac{1}{2} \lim_{n \to \infty} \int_{x_I=-\frac{1}{4}}^{x_I=\frac{1}{4}} \Delta_n(x_I^2 - (1/n)^2)\,dx_I\, 2^3 \{n_I(\boldsymbol{s}_a+1) - \boldsymbol{d}_{0,n_I}\} = \boldsymbol{s}_a \qquad 25$$

Here we note for completeness that, $m_I$ is summed over the set $\{-1,1\}$. In a similar way we find that for $\iota_3=1$ we have, $(s_{3I})_I=\sigma_a$. This leaves the term, $s_{2I}$, to be evaluated. In this case, integration over, $\mu_{2I}$, leads to $\sigma_a$. Because, there is no $n_I$ dependence in $s_{2I}$ the related sum term leads to unity. Also the sum over $m_I$ and the integration over $x_I$ cancel each other to unity, because there is also no, $\text{sign}[T\Delta_\varepsilon\text{-}\tau_I]$, term in $s_{2I}$. Hence, $(s_{2I})_I=\sigma_a$. This leads to $(A)_I=\sigma_a$, hence, because of symmetry of the standard normal density, $(A)=0$, because, $(1)_{II}=1$. Similarly, we find, $(B)=0$. Secondly, we aim to evaluate the variance $(A^2)$. Here, we also need to evaluate the three s-functions, because, the $\iota_j$, $j=1,2,3$, refer to exclusive sets. We note

$$A^2 = \sum_{k=1}^{3} \boldsymbol{i}_k(\boldsymbol{s}_a)\, s_{kI}^2(\boldsymbol{s}_a) \qquad 26$$

As in the previous case, we first evaluate $(s_{1I}^2)_I$. Because of the squaring, the $\text{sign}[T\Delta_\varepsilon\text{-}\tau_I]$, term in $s_{1I}$ collapses to unity. Moreover, squaring the term $\{n_I(\sigma_a+1)-\delta_{0,n_I}\}$, leads to $\{n_I^2+\delta_{0,n_I}^2\}$, because, the term, $n_I$, times $\delta_{0,n_I}$ cancels to zero. Hence,

$$\sum_{m_I} \frac{1+m_I}{8} \sum_{n_I=0}^{1} \frac{1}{2} \int_{-1/4}^{1/4} d\,x_I\, 2^3 \{n_I^2 + \boldsymbol{d}_{0,n_I}^2\} = 1. \qquad 27$$

Here we note for completeness that, $m_I$ is summed over the set $\{-1,1\}$. Hence, $(s_{1I}^2)_I=1$. Similarly, we find that $(s_{3I}^2)_I=1$. In addition, because, $s_{2I}^2=1$, we have to evaluate, $(1)_I$. Hence, $(s_{2I}^2)_I=1$. This leads us to the conclusion that $(A^2)=1$. In a similar way we can evaluate, $(B^2)=1$. The third and final step in the computation of the quantum correlation from local hidden variables, we have note that the two systems with index I and with index II are completely separated.

Hence, for (AB), we have to evaluate, $((A)_I(B)_{II})_N$. From the previous we saw that, $(A)_I=\sigma_a$, and, $(B)_{II}=-\sigma_b$. This then leads us to $(AB)=((A)_I(B)_{II})_N=-(\sigma_a\sigma_b)_N$. From eq.(20) we then may see that, the covariance term $(AB)=-(a_1b_1+a_2b_2+a_3b_3)$, leads to the expression

$$P(a,b) = \frac{([A-(A)][B-(B)])}{\sqrt{([A-(A)]^2)}\sqrt{([B-(B)]^2)}} = -\sum_{k=1}^{3} a_k\, b_k \,, \qquad 28$$

which is the quantum correlation. Note that the ( ) brackets represent the integrations, while the [ ] are used to keep the symbols 'together'.

*V. Conclusion and discussion.*

In the paper it was demonstrated that the quantum correlation can be obtained from a local hidden variables model. Unlike previous models of the author, this model has no 'strange' probability density function. In this case, the reason for the violation lies in the fact that a mathematical singularity can be hidden inside a sign function which is a valid element of the measurement function. Physically we could interpret this type of model by saying that here the blame for the violation is laid on the side of the measuring instruments, not on the side of a strange intermediate of local hidden variables. In the present model the Normal distributed variables take care of the binding together of the two separate measurement results. In the view of the author this is within the realm of orthodox classical probability theory and, hence, in the realm of classical physics.

Possible objections against the hiding of a singularity inside a sign function can alternatively be countered by noting that the usual Dirac delta contraction of an ordinary Normal density, is not the only possible expression for delta-function behaviour. In that respect, we may note the following definition of an alternative delta function as

$$\boldsymbol{d}_{Alt}(x) = \lim_{N\to\infty} \exp\left[-Nx - \frac{e^{-Nx}}{N}\right]. \qquad 29$$

This function can take values of 1, when x=0 and zero, when x≠0. This delta function is based on the following theta, or Heaviside function, as a derivative of the following theta,

$$\boldsymbol{q}_{Alt}(x) = \lim_{N\to\infty} \exp\left[-\frac{e^{-Nx}}{N}\right], \qquad 30$$

which is unity when, x≥0, while it vanishes when x<0. It can be verified rather easily that the results of the previous computations remain unaltered when T is replaced by 1 and the $\Delta_n$ functions are written as

$$\Delta_n(x_I^2 - c_e^2) \to \boldsymbol{d}_{Alt}(x_I),$$

$$\Delta_n(x_{II}^2 - c_e^2) \to \boldsymbol{d}_{Alt}(x_{II}). \qquad 31$$

Hence, the conclusions of the model remain the same, irrespective of the particular form of 'hidden' singularity one prefers to use.

Concerning Einstein's demand of realism and locality, it can be concluded that both are obeyed in the present model. Concerning the 'physical signals not faster than light' demand, the following remark is made.

As is well known, the axiom 'physical signals do not travel faster than light' entails counter-intuitive phenomena like FitzGerald-Lorentz contraction for moving objects and objects at rest. Naturally, the axiom must be valid in all kinds of physical situations, hence, also in e.g. a relativistic ideal gas. Suppose we inspect a situation for an arbitrary observer, $O_k$. In the view of $O_k$ we have

$$L_k = L_{kj} \sqrt{1 - \mathbf{b}^2_{kj}}.$$

Here, $L_k$ is the unit of length at rest in $O_k$'s frame of reference, while, $\beta_{kj}=v_{kj}/c$, is the (normed) velocity of another observer $O_j$'s frame relative $O_k$'s frame and $L_{kj}$ is the unit of length in motion, relative to observer $O_k$ and carried by observer $O_j$.

Introducing more than two observers, $O_k$ and $O_j$, we may introduce observer $O_i$ and note that, $L_k=L_{ki}(1-\beta^2_{ki})^{1/2}$. If, furthermore, $1 \geq \beta_{ki} > \beta_{kj} \geq 0$, we may write for $L_{ki}$ and $L_{kj}$

$$L_{kj} = L_{ki} \sqrt{1 - \mathbf{b}^2_{ki;j}}$$

with $\beta^2_{ki;j}$ denoting the 'velocity'

$$\mathbf{b}^2_{ki;j} = \frac{\mathbf{b}^2_{ki} - \mathbf{b}^2_{kj}}{1 - \mathbf{b}^2_{kj}}.$$

Note that, $1-\beta^2_{kj} \geq \beta^2_{ki}-\beta^2_{kj}$, hence, $\beta^2_{ki;j}$ in [0,1].

In addition to the $O_k$ 'point of view' we may also employ the $O_j$ 'point of view'. In this case we write, $L_j=L_{jk}(1-\beta^2_{jk})^{1/2}$, where $L_j$ is the unit of length at rest in $O_j$'s frame, $\beta_{jk}$ the velocity of $O_j$ relative $O_k$ and, $L_{jk}$ the 'currency' of $O_j$ when he wants to measure in $O_k$'s unit.

Hence the theorem:

*Because, $\mathbf{b}_{jk}=\mathbf{b}_{kj}$, we only may have, $L_{jk}=L_{kj}$, when, $L_j=L_k$.*

Subsequently, introducing more observers, $O_p$, $O_q$ and $O_r$ (particles in the gas) and performing a similar analysis, we may arrive at the following (Geurdes, (2001)). Firstly, when, $L_{pr}=L_{jk}$, from

$L_p=L_{jk}(1-\beta^2_{pr})^{1/2}$, it follows that

$$\beta^2_{pr} = 1 - ( L_p / L_{jk} )^2$$

Secondly, from, $L_{pq}=L_{ki}$, the expression $L_p=L_{ki}(1-\beta^2_{pq})^{1/2}$, leads to

$$\beta^2_{pq} = 1 - ( L_p / L_{ki} )^2$$

Let us furthermore state that, $L_j$ unequal to $L_k$. From the previous two equations for the 'compound velocities' $\beta_{pq;r}$ and $\beta_{ki;j}$ we find that,

$$\beta^2_{ki;j} = 1 - ( L_{kj} / L_{ki} )^2$$

together with

$$\beta^2_{pq;r} = 1 - ( L_{jk} / L_{ki} )^2$$

such that when, $\beta_{pq;r}=\beta_{ki;j}$, it easily follows that, $L_{jk}=L_{kj}$, despite $L_j$ unequal to $L_k$.

As a numerical example we may have,

$$\beta_{pr} = \sqrt{3/5}, \quad \beta_{pq} = \sqrt{3/2} \quad \beta_{pr} = \sqrt{9/10},$$
$$\beta_{kj} = \sqrt{1/3}, \quad \beta_{ki} = \sqrt{5/2} \quad \beta_{kj} = \sqrt{5/6}.$$

39

This example shows that the contradiction may have a physical basis. Moreover, for our purposes, it casts doubt upon Einstein's third demand that no physical signals may travel faster than light. If the FitzGerald-Lorentz contraction can turn into a contradiction in a serious physical environment such as a relativistic ideal gas, then it is not unthinkable that its basic principle, 'physical signals do not travel faster than light' is also not always valid.

Concludingly, a counter-example to Bell's theorem is found and the premises on which such theorem is based were discussed. Both theorem as well as its premises turned out to be disputable.